\documentclass[showpacs,preprintnumbers,showkeys,amsmath,amssymb]{revtex4}
\usepackage{graphicx}
\usepackage{dcolumn}
\usepackage{bm}
\begin{document}

\preprint{}

\title{Stability and phase transition of localized modes in Bose-Einstein condensates with both two- and three-body interactions\footnote{Published in Annals of Physics \textbf{360}, 679--693 (2015)}}

\author{Xiao-Dong Bai, Qing Ai, Mei Zhang,  Jun Xiong,\footnote{Corresponding author:
junxiong@bnu.edu.cn} Guo-Jian Yang,  and  Fu-Guo Deng}

\address{Department of Physics, Applied Optics Beijing Area Major Laboratory, Beijing normal University, Beijing 100875, China}

\begin{abstract}
We investigate the stability and phase transition of localized modes
in Bose-Einstein Condensates (BECs) in an optical lattice with  the
discrete nonlinear Schr\"{o}dinger model by considering  both  two-
and three-body interactions. We find that there are three types of
localized modes,  bright discrete breather (DB), discrete kink (DK),
and  multi-breather (MUB).  Moreover,  both two- and three-body
on-site repulsive interactions can stabilize DB, while on-site
attractive three-body interactions  destabilize it. There is a
critical value for the three-body interaction  with which both DK
and MUB become the most stable ones. We give analytically the energy
thresholds  for the destabilization of localized states and find
that they are unstable (stable) when the total energy of the system
is higher (lower) than the thresholds. The stability and dynamics
characters of DB and MUB are general for extended lattice systems.
Our result is useful for the blocking, filtering, and transfer of
the norm in nonlinear lattices for BECs with both two- and
three-body interactions.
\end{abstract}



\maketitle

\section{Introduction}

Localized excitation in Bose-Einstein condensates (BECs) has
become one of the most interesting topics in nonlinear lattice
systems since the discrete breathers (DBs) and the intrinsic
localized modes are discovered
\cite{YIshimori1982,MPeyrard1984,Sievers1988}. The most well-known
fascinating feature of the localized mode is that it can propagate
without changing its shape as a result of the balance between
nonlinearity and dispersion
\cite{DKCampbell2004,Campbell2004,SFlach2008,HHennig2013}. DB
arises intrinsically from the interplay between the nonlinearity
and the discreteness of the system. DB has been observed in
various systems, such as micromechanical cantilever arrays
\cite{MSato2006}, antiferromagnet systems
\cite{UTSchwarz1999,MSato2004}, Josephson-junction arrays
\cite{ETrias2000,AVUstinov2003}, nonlinear waveguide arrays
\cite{RMorandotti1999,HSEisenberg1998}, BECs
\cite{BEiermann2004,ATrombettoni2001}, Tonks gas
\cite{Kolomeisky1992}, superfluid fermi gases \cite{JuKuiXue2008},
and some dissipative systems \cite{MGClerc2011,PCMatthews2011}.
The static, dynamical, and other properties of DB have been
studied theoretically in the last decade
\cite{SAubry1997,JDorignac2004,HSakaguchi2010,MMatuszewski2005,NBoechler2010}.
It has been demonstrated that DBs are attractors in dissipative
systems \cite{SFlach1998,PJMartinez2003,RSMacKay1998}, or act as
virtual bottlenecks which slow down the relaxation processes in
generic nonlinear lattices
\cite{GPTsironis1996,RLivi2006,GSNg2009}. It is shown that the
stability of the discrete localized modes plays a crucial role in
blocking, filtering, and transfer of the norm through a localized
mode. By far, there are many interesting works focused on this
stability by considering two-body interactions
\cite{HHennig2010,RFranzosi2011,LuLi2005,ZXLiang2005,yingwu2004,XDBai2012}.

Recently, the three-body interactions could be observed or
realized in experiment and theory \cite{SWill2010,DaleyAJ2014}. In
2014, Petrov \cite {WSBakr2014}  proposed a method to control the
two- and three-body interactions in  ultracold Bose gas in any
dimension. The three-body interactions play an important role in
many interesting physical phenomena
\cite{TLuu2007,BolunChen2008,PRJohnson2009,KezhaoZhou2010}, and
even lead to a variety of unique properties that are absent in the
system dominated by the two-body interactions which can be
governed by a Feshbach resonance \cite{SInouye1998}. For example,
in 2010, Dasgupta \cite{RDasgupta2010} discovered that if the
two-body interactions are attractive, the presence of three-body
interactions makes the crossover process from
Bardeen-Cooper-Schrieffer (BCS) to Bose-Einstein condensates
(BECs) a nonreversible one. In 2012, Singh \emph{et al.}
\cite{MSingh2012} found that the coupling of the two- and
three-body interactions can affect strongly  the transition from
Mott insulator to superfluid for ultracold bosonic atoms in an
optical lattice or a superlattice. Up to now, there are few works
on localized excitations in nonlinear lattice systems by
considering three-body interactions. Especially, there are no
systematical analysis of the types,  existence, and stability of
the localized modes, such as DB, the discrete kink (DK)
\cite{PGKevrekidis2002,YuAKosevich2004,JMSpeight1999}, and
multi-breather (MUB). It is natural to ask how the two- and
three-body interactions affect these properties of the localized
modes in BECs.

In this paper, we investigate the stability and phase transition of
localized modes in  BECs in an optical lattice with a discrete
nonlinear Schr\"{o}dinger model (DNLS) in the case by considering
both two- and three-body interactions. We find that there are three
different types of localized modes, that is, DB, DK, and MUB, and
give the critical conditions for these localized modes. Both the
two- and three-body on-site repulsive interactions can stabilize DB,
while the three-body on-site attractive interactions destabilize it.
We calculate analytically the energy thresholds, the Peierls-Nabarro
(PN) energy barrier \cite{YSKivshar1993,BRumpf2004}, characterizing
the stability of the localized excitation modes.  If the total
energy of the system is higher (lower) than the  thresholds, the
localized states are unstable (stable). Moreover, the stability and
dynamics characters of DB and MUB are general for extended lattice
systems.  Our result is important for the transfer of BECs through
the discrete localized modes, and is useful for controlling the
transmission of matter waves in interferometry and
quantum-information processes when there are both two- and
three-body interactions in the system.

\section{localized states and  Peierls-Nabarro barrier}\label{sec2}

\subsection{The model}\label{sec21}

Besides the on-site two-body interactions, let us investigate the
effect on transfer of BECs through discrete localized mode from
the on-site three-body interactions of ultracold Bose gas in an
optical lattice.  Under the mean-field theory, the Hamiltonian of
the system of BECs in an optical lattice can be written as
\cite{AXZhang2008}:
\begin{eqnarray}\label{eq1}
\! \mathcal{H}\!\!=\!\!\sum _{n=1}^M\!\frac{U_1}{2}|\psi
_n|^4
 \!+\!\sum_{n=1}^{M}\!\frac{U_2}{3}|\psi _{n}|^6\!-\!\frac{J}{2}\!\sum_{n=1}^{M-1}\! \left(\psi _n^*\psi
 _{n+1}\!+c.c\right).
\end{eqnarray}
Here $n$ $(=1,\cdots,M)$ is the index of the site. $\psi_n$ is a
complex variable and $|\psi_n(t)|^2\equiv N_n(t )$ is the mean
number of bosons at site $n$ (i.e., the norm $N_n(t)$). The first
two terms represent the mean-field two- and three-body interaction
energy, respectively, and the third term describes the hopping
between nearest-neighboring sites. $U _1=4\pi\hbar^2a_s V_{eff}/m$
represents the effective on-site inter-atomic two-body
interaction, where $V_{eff}$ is the effective mode volume of each
site, $m$ is the atomic mass, and $a_s$ is the s-wave atomic
scattering length. Here, we focus on the repulsive two-body
interaction, i.e., $U_1>0$. $U_2$ represents the effective on-site
inter-atomic three-body interactions, including both the repulsive
and the attractive three-body interactions which are represented
by $U_2>0$ and $U_2<0$, respectively. $J$ is the tunneling
amplitude. Within the canonical equation $i\frac{\partial \psi _n}
{\partial \tau}=\frac{\partial H}{\partial \psi_n^*}$, one can
obtain the dimensionless DNLS
\cite{AXZhang2008,BBBaizakov2009,EWamba2011}
\begin{eqnarray}\label{eq2}
\!i\frac{\partial \psi _n} {\partial t}\!=\!\lambda_1 \left|\psi
_n\right|{}^{2}\psi_n \!+ \!\lambda_2\left|\psi
_n\right|{}^{4}\psi_n \!-\!\frac{1}{2}\left[\psi _{n-1}\!+\psi
_{n+1}\right],
\end{eqnarray}
where
\begin{eqnarray}\label{eq3}
\sum_{n=1}^M|\psi_n|^2=1.
\end{eqnarray}
Here, $\lambda_1=U_1/J$, $\lambda_2=U_2/J$ and $t=J\tau$ are the
normalized dimensionless two-body interaction, three-body
interaction and time, respectively. Assume that
$\psi_n(t)=A_n(t)\text{exp}\left(\text{$i\theta $}_n(t)\right)$,
the Hamiltonian $H$ becomes
\begin{eqnarray}\label{eq1a}
 H\!=\!\!\sum _{n=1}^M\!\!\left(\frac{\lambda_1}{2}A_n^4\!+\!\frac{\lambda_2}{3}A_n^6\right)
\!\!-\!\!\!\sum_{n=1}^{M-1}\!\! \left[A_n
A_{n+1}\!\cos(\theta_n\!\!-\!\theta_{n+1})\right].
\end{eqnarray}

Usually, we use the Peierls-Nabarro (PN) energy landscape
\cite{YSKivshar1993,BRumpf2004} to reflect the fact that
discreteness breaks the continuous translational invariance of a
continuum model. It is related to the PN potential whose amplitude
can  be seen as the minimum barrier which should be overcome to
translate an object by one site.  As in Ref.
\cite{HHennig2010},  the PN energy landscape is defined as follows:
for a given  configuration  of  amplitudes  $A_n$,  with  respect to
the phase  difference  $\delta\theta_{ij}= \theta_i - \theta_j$, the
PN energy landscape is obtained by extremizing $H$
\begin{equation}\label{eq4}
H_{\text{PN}}^l=\underset{\delta\theta _{ij}}{\min }(-H), \;\;\;\;\;\;\;\;\text {
}H_{\text{PN}}^u=\underset{\delta \theta _{ij}}{\max }(-H),
\end{equation}
where $H_{\text{PN}}^l$ and $H_{\text{PN}}^u$ are the lower and
upper parts of the PN landscape, respectively.

In order to give an insight into the dynamical behavior of BECs in
an optical lattice with both two- and three-body interactions,  we
mainly consider the nonlinear trimer model, i.e.,  the DNLS with
$M=3$ lattice sites. In this case, the
 Hamiltonian $H$ of the system is
\begin{eqnarray}\label{eq5}
\!\!\!\!\!\!H_{M=3}\!\!&\!=\!&\!\!\frac{\lambda_1
}{2}\left(A_1^{4}+A_2^{4}+A_3^{4}\right)+\frac{\lambda_2
}{3}\left(A_1^{6}+A_2^{6}+A_3^{6}\right) \nonumber\\&&
-\left[A_1A_2\cos \left(\theta_1\!\!-\theta
_2\right)\!+\!A_2A_3\cos \left(\theta _2\!\!-\theta
_3\right)\right],
\end{eqnarray}
where $\delta \theta _{12},\delta \theta _{23}\in [0,\pi ]$.
When $\delta\theta_{12}=\delta\theta_{23}=0$, one can get
the upper PN energy landscape
\begin{eqnarray}\label{eq6}
H_{\text{PN}}^u\!\!&=&\!\!-\frac{\lambda_1
}{2}\left(A_1^{4}+A_2^{4}+A_3^{4}\right) -\frac{\lambda_2
}{3}\left(A_1^{6}+A_2^{6}+A_3^{6}\right)\;\;\;\;\nonumber\\&&+\left(A_1+A_3\right)A_2.
\end{eqnarray}
When $\delta\theta_{12}=\delta\theta_{23}=\pi$, the lower PN
energy landscape can be obtained as
\begin{eqnarray}\label{eq7}
H_{\text{PN}}^l\!\!&=&\!\!-\frac{\lambda_1
}{2}\left(A_1^{4}+A_2^{4}+A_3^{4}\right) -\frac{\lambda_2
}{3}\left(A_1^{6}+A_2^{6}+A_3^{6}\right)\;\;\;\;\nonumber\\&&-\left(A_1+A_3\right)A_2.
\end{eqnarray}
The lower and the upper parts of the PN landscape bound the phase
space of the trimer \cite{HHennig2010}. Because the localized mode whose properties we are studying
corresponds to the minimum on $ H_{\text{PN}}^{l}$, we should focus
on the lower PN landscape,  i.e.,   $\delta \theta
_{12}=\delta \theta _{23}=\pi$.

\subsection{Phase transition of localized states and  Peierls-Nabarro barrier}\label{sec22}

To investigate the type  of localized modes and its stability when
the norm transfers through a localized mode, 
we use the nonlinear trimer model ($M = 3$ for Eq.(\ref{eq2}))
\begin{equation}\label{eq8}
\begin{split}
 i\partial _t \psi _1&=\lambda_1 \left|\psi
_1\right|{}^{2}\psi _1+\lambda_2\left|\psi _1\right|{}^{4}\psi
_1-\frac{1}{2}\psi _2,\\
 i\partial _t \psi _2&=\lambda_1 \left|\psi
_2\right|{}^{2}\psi _2+\lambda_2\left|\psi _2\right|{}^{4}\psi
_2-\frac{1}{2}\left(\psi _1+\psi _3\right),\\
 i\partial _t \psi _3&=\lambda_1 \left|\psi
_3\right|{}^{2}\psi _3+\lambda_2\left|\psi _3\right|{}^{4}\psi
_3-\frac{1}{2}\psi _2.
\end{split}
\end{equation}
Here, the normalization reads $ N=\sum _{n=1}^3 \left|\psi
_n\right|{}^2=1$. By setting $\delta \theta _{12}=\delta \theta
_{23}=\pi$, from Eq. (\ref{eq8}) one can get
\begin{equation}\label{eq9}
\begin{split}
&2\lambda_1 A_1 \left(1-2 A_1^2-A_3^2\right)+2 \lambda_2 A_1
\left(1-A_3^2\right)\!\!\left(1-2
A_1^2-A_3^2\right)\\
&-\sqrt{1-A_1^2-A_3^2}+\frac{A_1\left(A_1+A_3\right)}{\sqrt{1-A_1^2-A_3^2}}=0, \\
 &2 \lambda _1A_3 \left(1-2 A_3^2-A_1^2\right)
 +2 \lambda _2A_3 \left(1-A_1^2\right)\!\!\left(1-2 A_3^2-A_1^2\right)
 \\&-\sqrt{1-A_1^2-A_3^2}+\frac{A_3\left(A_1+A_3\right)}{\sqrt{1-A_1^2-A_3^2}}=0.
\end{split}
\end{equation}
Let us define
\begin{eqnarray}\label{eq10}
\frac{\partial^2H_{\text{PN}}^l}{\partial A_1^2}\equiv H_1,\;\;\;\;
\frac{\partial^2H_{\text{PN}}^l}{\partial A_1 \partial A_3}\equiv
H_2,\;\;\;\;    \frac{\partial^2H_{\text{PN}}^l}{\partial
A_3^2}\equiv H_3,\;\;
\end{eqnarray}
and give inequality
\begin{eqnarray}\label{eq11}
H_2^2-H_1H_3\geq0.
\end{eqnarray}
By solving Eq. (\ref{eq9}) one can get the different kinds of
solutions,  including the stationary states and the saddle points.
By substituting the saddle points obtained from Eq. (\ref{eq9}) into
Eq. (\ref{eq11}), one can get the critical conditions which are the
boundaries between phases I-III in Fig. 1. Actually, the
boundaries can also be gained numerically from Eq. (\ref{eq9}).

The results show that phase I occurs when $\lambda_1$ and
$\lambda_2$ satisfy the relation
\begin{eqnarray}\label{eq12}
\lambda _2>1.68708-1.08801\lambda _1+0.00522\lambda _1^2.
\end{eqnarray}
For this case the norm can be pinned in  any one of the three
sites, shown in Fig. 2(a)-(c).  There  exists  DB. That
is,  the norm is pinned in the middle site.

Phase III appears when $\lambda_1>3.5$ and $\lambda_1$ and
$\lambda_2$ satisfy the relation
\begin{eqnarray}\label{eq13}
\begin{split}
\lambda_l<\lambda_2 <\lambda_u,
\end{split}
\end{eqnarray}
where
\begin{eqnarray}\label{eq14}
\begin{split}
\lambda_l&=7.39452-4.33978\lambda
_1+0.37649\lambda _1^2-0.0178\lambda _1^3,\;\;\;\;\;  \\
\lambda _{u}&=-12.223+6.6531\lambda _1-1.722\lambda
_1^2+0.1715\lambda _1^3.
\end{split}
\end{eqnarray}
In this case, the norms are localized in the adjacent two sites,
shown in Fig. 2(e)-(g). It is called as DK.

Phase II is the part other than phases I and III, shown in Fig. 1.
For this case, the norms of the three sites are nearly equivalent
and there is no saddle point on the contour plots of the lower PN
energy landscape $H_{PN}^l$, shown in  Fig. 2(d) and (h). In this
case, no localized mode exists.

To investigate the stability of the localized states, we pay our
attention to the PN barrier. As shown in Fig. 2, the
projection of $H_{PN}^{l}$ onto the $A_1-A_3$ plane exhibits one,
two, or three minima. Each minimum refers to a  stationary  state.
If there exist saddle points, the localized mode  appears and these
stationary states correspond to DB or DK. It is clear that the
existence of DB, DK, and saddle points depends strongly on
$\lambda_1$ and $\lambda_2$. We define the energy of  DB (DK) as
$E_{DB}$ ($E_{DK}$),  the energy of the saddle points as $E_{thr}$,
and the energy difference $\Delta E = E_{thr}-E_{DB(DK)}$. Actually,
$\Delta E$ is  the PN  barrier. When the total energy of the trimer
$E_t=-H>E_{thr}$ or $E_t-E_{DB(DK)}>E_{thr}-E_{DB(DK)}=\Delta E$,
the DB (DK) should be unstable. On the contrary, when
$E_t-E_{DB(DK)}<\Delta E$, the DB (DK) should be stable. We should
note that $E_{thr}$ merely marks the energy of the trimer at the
saddle point and is identified with the destabilization threshold of
the DB (DK). $\Delta E$ is the energy difference between the
stationary state and the saddle points. Hence, $\Delta E$ represents
the minimum energy barrier required to translate  BECs  by one
lattice site. The higher  $\Delta E$ is, the more stable the DB (DK)
is. Therefore, the quantity $\Delta E$ can provide a deep insight
into the stability property of DB (DK) for different values
$\lambda_1$ and $\lambda_2$.

\begin{figure}   
\includegraphics[width=8cm]{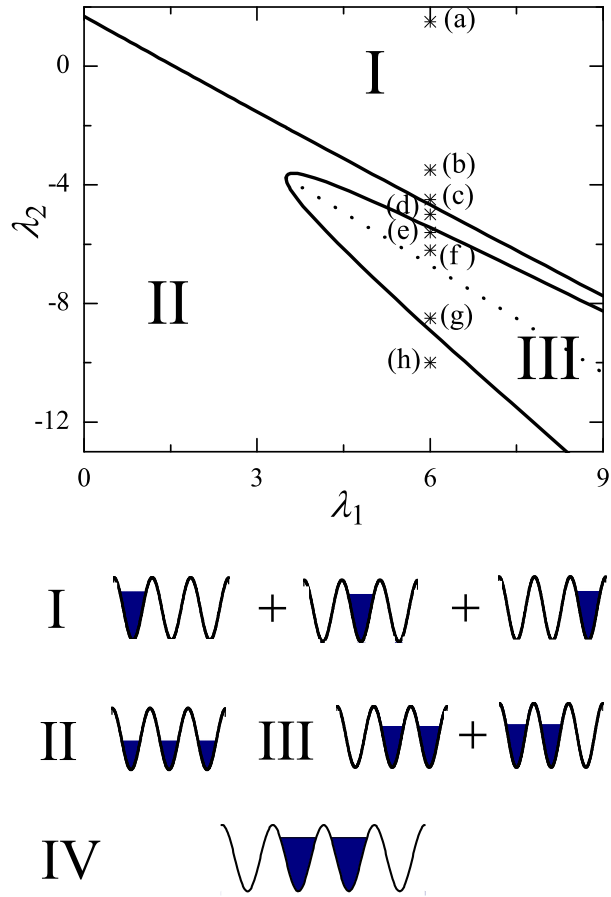}
\caption{ \label{Fig1} (Color online) The norm distribution and
the structure of the localized states with  two-body interactions
($\lambda_1$) and three-body interactions ($\lambda_2$). The solid
lines are the boundaries dividing phases I-III. The dotted line
presents the critical value $\lambda_2^*$ with $\lambda_1$. When
$\lambda_2=\lambda_2^*$ the DK is the most stable one. The symbols
(a)-(h) correspond to the sets of parameters
$(\lambda_1,\lambda_2)$ used in Fig. 2. Phase IV is gained by Eq.
(\ref{eq16}) which corresponds to a four-site model.}
\end{figure}
\begin{figure*}   
\includegraphics[width=16cm]{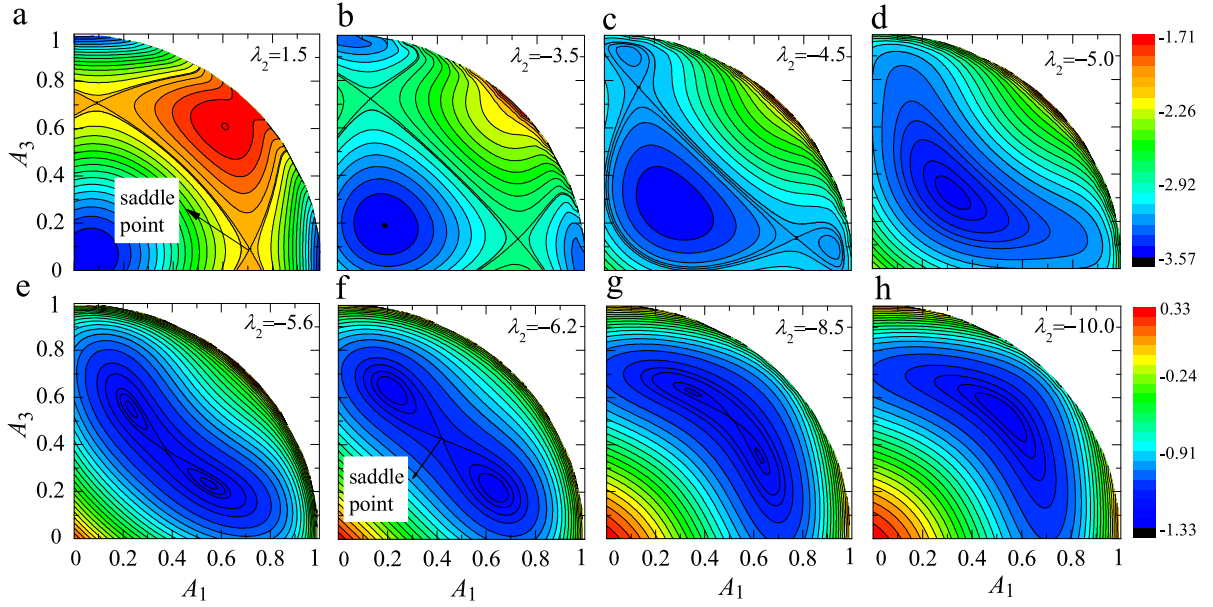}
\caption{ \label{Fig2} (Color online) Contour plots of the lower
PN energy landscape $H_{PN}^l$ for different $\lambda_2$ with
fixed $\lambda_1=6$. (a)-(h) correspond to the points a-f marked
in Fig. 1. In (a)-(c), the three minima are separated by the two
saddle points and DB exists. In (d) and (h), no saddle point
exists, which means there is no any kind of localized modes. In
(e)-(g), two minima are separated by one saddle point and DK
exists. The color codes present the energy of $H_{PN}^l$.}
\end{figure*}

The structure of MUB in the extended lattices is similar to that of
a DK. MUB is a four-site solution of the DNLS, with high atomic
density concentrated mainly at two middle sites and two low-density
sites. Up to now, there are few  studies on the properties of  MUB,
especially on its stability. Here we  also investigate the effect on
transfer of BECs through the MUB from the on-site two- and
three-body interactions. It can be investigated by using four-site
model, i.e., $M=4$.  Eq. (\ref{eq8}) becomes
\begin{equation}\label{eq15}
\begin{split}
 i\partial _t \psi _1&=\lambda_1 \left|\psi
_1\right|{}^{2}\psi _1+\lambda_2\left|\psi _1\right|{}^{4}\psi
_1-\frac{1}{2}\psi _2,\\
 i\partial _t \psi _2&=\lambda_1 \left|\psi
_2\right|{}^{2}\psi _2+\lambda_2\left|\psi _2\right|{}^{4}\psi
_2-\frac{1}{2}\left(\psi _1+\psi _3\right),\\
 i\partial _t \psi _3&=\lambda_1 \left|\psi
_3\right|{}^{2}\psi _3+\lambda_2\left|\psi _3\right|{}^{4}\psi
_3-\frac{1}{2}\left(\psi _2+\psi _4\right), \;\;\\
i\partial _t \psi
_4&=\lambda_1 \left|\psi _4\right|{}^{2}\psi
_4+\lambda_2\left|\psi _4\right|{}^{4}\psi _4-\frac{1}{2}\psi _3.
\end{split}
\end{equation}
Here, the normalization reads $ N=\sum _{n=1}^4 \left|\psi
_n\right|{}^2=1$. Similarly, by setting $\delta \theta _{12}=\delta
\theta _{23}=\delta \theta _{34}=\pi$,  one can get
\begin{equation}\label{eq16}
\begin{split}
\lambda_1\!\!\left(A_1^2\!-\!A_2^2\right)\!&+\!\lambda_2 \!\!\left(A_1^4\!-\!A_2^4\right)\!+\!\frac{1}{2}\!\left(\frac{A_2}{A_1}\!-\!\frac{A_1+A_3}{A_2}\right)\!\!=0,\\
\lambda_1\!\!\left(A_2^2\!-\!A_3^2\right)\!&+\!\lambda_2\!\!\left(A_2^4\!-\!A_3^4\right)\!+\!\frac{1}{2}\!\left(\frac{A_1+A_3}{A_2}\!-\!\frac{A_2+A_4}{A_3}\right)\!\!=0,\\
\lambda_1\!\!\left(A_3^2\!-\!A_4^2\right)\!&+\!\lambda_2\!\!\left(A_3^4\!-\!A_4^4\right)\!+\!\frac{1}{2}\!\left(\frac{A_2+A_4}{A_3}\!-\!\frac{A_3}{A_4}\right)\!\!=0.
\end{split}
\end{equation}
By solving Eq. (\ref{eq16}), one can get the MUB and the saddle
points.

Now, let us discuss the stability and dynamics of DB and DK on
phases I and III in which the system has saddle points (i.e., the
system has the Peierls-Nabarro barriers), excluding phase II in
which there is only one stationary state without saddle points. As
shown in previous works \cite{BRumpf2004,HHennig2013}, if there is
no saddle point, the system should be in a random (generic) state in
the presence of boundary or other local dissipation, and cannot form
any kind of localized modes. That is, for the system with parameters
of phase II, the localized mode does not occur. Although phase IV
cannot be shown in Fig. 2 as there are not enough
dimensions,  saddle points can still exist and they are also
investigated here.

\section{Stability of localized states}\label{sec3}

The contours of the lower PN energy landscape $H_{PN}^l$ for
different $\lambda_2$ with fixed $\lambda_1=6$ are shown in Fig.
2(a)-(h). Fig. 2(a)-(c) corresponds  to phase I in Fig. 1 and
there exist three stationary points and two saddle points. That
is, the norm can be localized in three different ways (see Fig. 1)
and DB exists.  Fig. 2(e)-(g) corresponds to phase III in Fig. 1
and there exist two stationary points and one saddle point [this
case cannot exist in the system without three-body interactions,
as shown in Fig. 1 with $\lambda_2=0$]. This case corresponds to
DK.  Fig. 2(d) and (h) corresponds  to phase II in Fig. 1 and
there exists only one stationary point but no saddle point.

\begin{figure*}   
\includegraphics[width=16cm]{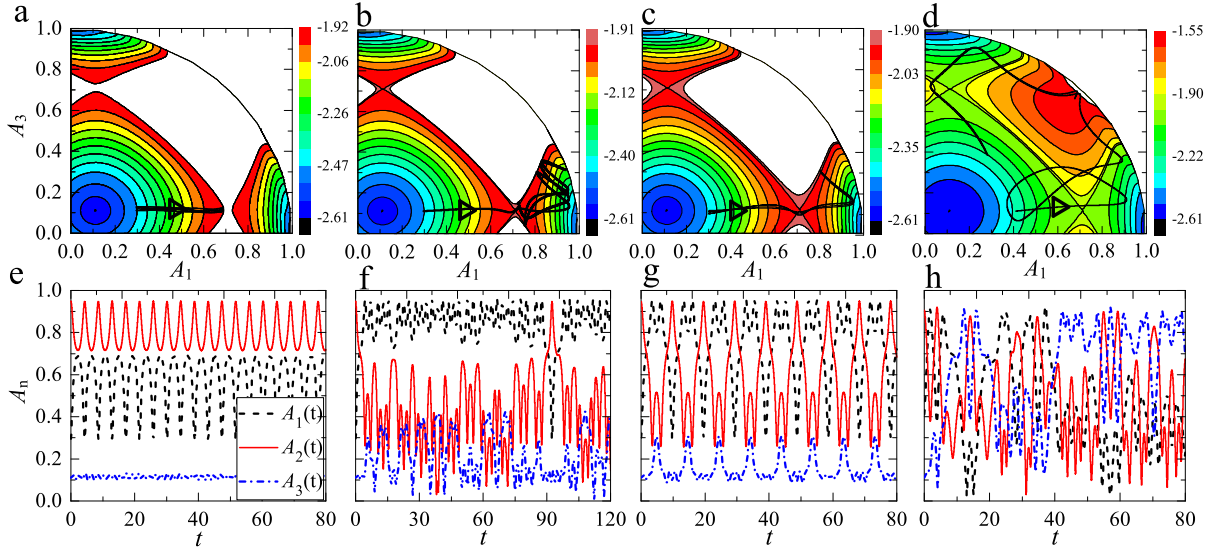}
\caption{\label{Fig3}  (Color online) Dynamics of the trimer when
its total energy is increased with fixed two-body interactions
$\lambda_1=6$ and three-body interactions $\lambda_2=-1.5$, where
$E_{thr1}=-1.91469$. A projection of the orbit onto the $A_1-A_3$
plane is over-plotted (black curve) in (a)-(d). (a) and (e)
$E_t=-1.9176 < E_{thr1}$, areas in phase space are disconnected
and the amplitudes $A_n(t)$ indicate that the maximum amplitude
remains at  site 2, i.e., the DB is stable; (b) and (f)
$E_t=-1.91179>E_{thr1}$, areas in phase space are connected. A
slight instability of the DB centered at site 2 is observed, the
breather migrates to site 1, and then tangles in site 1; (c) and
(g) for a high total energy $E_t=-1.8995$, the norm is transmitted
to site 1, tangles,  and then comes back; (d) and (h) for an
enough  high  total energy $E_t=-1.5504$,  the orbit explores
large parts of the phase space and visits all three sites. In all
cases $\delta_{\theta}=\pi$.}
\end{figure*}

\subsection{The stability and dynamics of DB}\label{sec3.1}

To study the transfer of  BEC through
the DB in an optical lattice with three-body interactions, we
should pay attention to the energy threshold $E_{thr}$ and energy
$E_{DB}$ again. In this case, for $\lambda_1\rightarrow\infty$ and
$\lambda_2/ \lambda_1\rightarrow 0$, one can get the saddle point
from Eq. (\ref{eq9}) as
\begin{eqnarray}\label{eq17}
\begin{split}
A_1&\approx\frac{1}{\sqrt{2}}+\frac{\lambda  _2^2}{\sqrt{2}
\lambda
_1^3}+\frac{\lambda  _2}{2 \sqrt{2} \lambda _1^3},\\
A_3&\approx\frac{3}{2 \sqrt{2} \lambda _1^3}-\frac{1}{\sqrt{2}
\lambda _1^2}+\frac{1}{\sqrt{2}\lambda _1}-\frac{\lambda  _2}{2
\sqrt{2} \lambda _1^2}.
\end{split}
\end{eqnarray}
By substituting the saddle point into Eq. (\ref{eq7}) with
$A_1^2+A_2^2+A_3^2=1$, one can get the energy threshold
\begin{eqnarray}\label{eq18}
E_{\text{thr1}}\!\!&=&\!-\!\frac{\lambda
_1}{4}-\frac{1}{2}-\frac{1}{4 \lambda _1}+\frac{1}{4 \lambda
_1^2}-\!\!\frac{1}{4 \lambda _1^3}+\frac{9}{16 \lambda
_1^4}-\frac{3}{2 \lambda _1^5}
\nonumber\\
&&\!-\!\!\left(\frac{1}{12}-\frac{1}{8 \lambda _1^2}+\frac{1}{4\text{
}\lambda _1^3}-\frac{3}{8 \lambda _1^4}+\frac{5}{8 \lambda
_1^5}\right)\!\! \lambda _2\nonumber\\
&&\!-\!\!\left(\frac{1}{16
\lambda_1^3}\!-\!\frac{3}{16 \lambda _1^4}\!-\!\frac{1}{2 \lambda
_1^5}\right)\!\! \lambda _2^2+\!\!\left(\frac{1}{32 \lambda
_1^4}\!-\!\frac{5}{4 \lambda _1^5}\right)\!\! \lambda _2^3.\nonumber\\
\end{eqnarray}
Of course, $E_{DB}$ can also be obtained by substituting the
bright breather given by Eq. (\ref{eq9}) into Eq. (\ref{eq7}).

Next, we consider the fixed  point corresponding to the bright
breather which is gained from Eq. (\ref{eq9}). An initial condition
for the bright breather reads
$\overrightarrow{\psi}^{DB}(0)=(A_1^{DB},A_2^{DB},A_3^{DB})$. We add
perturbations  to site 1:
$\overrightarrow{\psi}{(t=0)}=((A_1^{DB}+\delta_
{A_1})e^{i\delta_{\theta}},A_2,A_3^{DB})$, where $A_2 =
(1-|\psi_1|^2-|\psi_3|^2)^{1/2}$. Compared to the bright breather,
we add an amplitude $\delta_{A_1}$ to site $1$ and rotate the phase
$\theta_1$ by $\delta_{\theta}$. Dynamics on the PN landscape for
increasing total energy of the trimer with fixed two-body
interactions $\lambda_1=6$ and three-body interactions
$\lambda_2=-1.5$ is shown in Fig. 3. Here
$E_{thr1}=-1.91469$, and  we  fix $\delta_{\theta}=\pi$ and increase
$\delta_{A_1}$ in Fig. 3(a)-(d). If the perturbation is
small, $E_t<E_{thr1}$, the areas in phase space are disconnected.
Furthermore, the DB is stable and practically no transfer of norm
takes place on short time scales [see Fig. 3(a) and (e)].
If the perturbation is large, $E_t>E_{thr1}$, the areas in phase
space are connected [see Fig. 3(b) and (c)].  Instability
of the DB centered at site 2 can be observed, that is, the breather
migrates to site 1 and norm is transferred to site 3 [see Fig.
3(b), (c), (f), and (g)]. If the perturbation is large
enough, the orbit  goes  out of the regular island into the chaotic
sea [see Fig. 2(d)] and large amplitudes $A_n(t)$ are found at all
the three sites, as depicted in Fig. 3(h).

\begin{figure}     
\includegraphics[width=12cm]{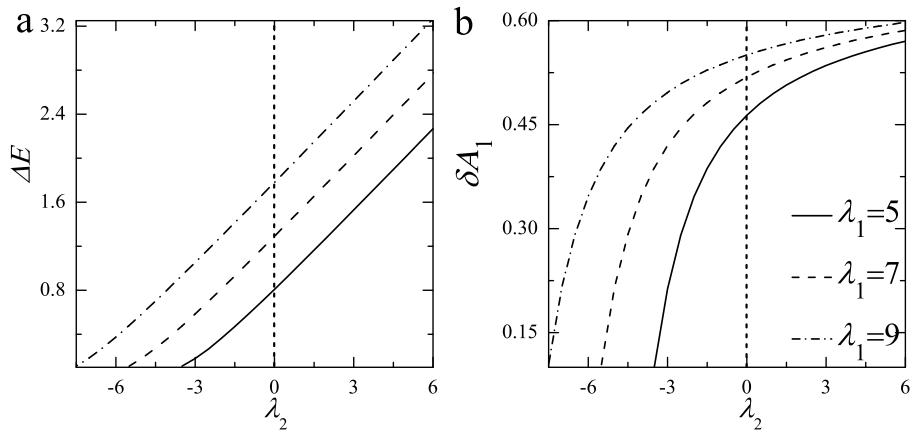}
\caption{\label{Fig4} (Color online) (a) Energy difference $\Delta
E$ and (b) perturbations $\delta A_1$ for destabilizing the DB  as
a function of three-body interactions $\lambda_2$. Different lines
indicate three different values of $\lambda_1$, respectively. The
left and right of the dotted line present attractive (i.e.,
$\lambda_2<0$) and repulsive (i.e., $\lambda_2>0$) three-body
interactions, respectively. $\delta \theta_{12}=\pi/4$ in (b).}
\end{figure}

Further, we use the parameters $\delta A_1$ and $\Delta E$ to
investigate the effect of the two- and three-body interactions on
the stability of DB, shown in Fig. 4. It is clear that both
$\Delta E$ and $\delta A_1$ increase with $\lambda_1$ whether
three-body interactions are repulsive or attractive (i.e.,
$\lambda_2>$ or $\lambda_2<0$), which means that a larger
perturbation is required to destabilize the DB when $\lambda_1$
increases and on-site two-body interactions can stabilize the DB.
Interestingly,  both $\Delta E$ and $\delta A_1$ increase with
repulsive three-body interactions ($\lambda_2>0$) but decrease with
attractive three-body interactions ($\lambda_2<0$), which means that
a relatively large perturbation is needed to destabilize the DB when
repulsive three-body interactions increase. When attractive
three-body interactions increase, a relatively small perturbation is
needed to destabilize the DB. That is, repulsive on-site three-body
interactions can stabilize the DB, while attractive on-site
three-body interactions destabilize the DB. For large enough
attractive on-site three-body interactions, $\Delta E=\delta A_1=0$
and the DB is completely unstable.

\begin{figure*}   
\includegraphics[width=16cm]{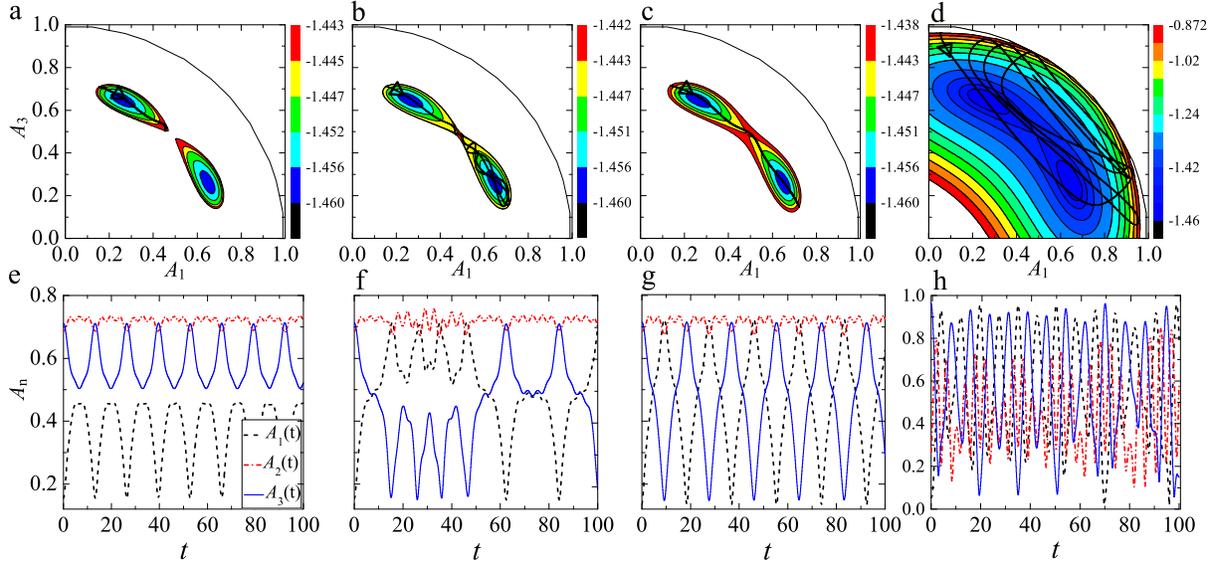}
\caption{\label{Fig5} (Color online) Dynamics of the trimer when
its total energy is increased  with fixed two-body interactions
$\lambda_1=6$ and three-body interactions $\lambda_2=-7.5$, where
$E_{thr2}=-1.44281$. A projection of the orbit onto the $A_1-A_3$
plane is over-plotted (black curve) in (a)-(d). (a) A contour plot
of the lower PN energy landscape $H_{PN}^l$ is shown for the total
energy of the trimer below the threshold ($E_t=-1.44304<
E_{thr2}$). Obviously, the areas in phase space are disconnected.
(e) The amplitudes $A_n(t)$ with time $t$ indicate that the
dominating amplitude remains at sites 2 and 3,  i.e., the norm is
still localized in  sites 2 and 3 and the DK is stable. (b) and
(f) for total energy $E_t=-1.44238>E_{thr2}$, areas in phase space
are connected. A slight instability of the DK centered at site 1
is observed. The DK migrates to site 1 and then tangled at site 1
or 3, but the norm of site 2 is nearly constant. (c) and (g) for a
high total energy $E_t=-1.43809>E_{thr2}$, the DK migrates to site
1 and then come back more easily. Long-range and long-lived
Josephson oscillations between sites 1 and 3 with negligible
variation of norm in site 2 are observed. i.e., the DK is unstable
and the dominating amplitude does not remain at  sites  2 and 3.
(d) and (h) for an enough  high total energy $E_t=-0.871835$, the
orbit is  out of the regular island into the chaotic sea, and the
dominating amplitude can be found in any of the three sites, which
means the DK is unstable. In all cases $\delta_{\theta}=0$.}
\end{figure*}

\begin{figure}     
\includegraphics[width=12cm]{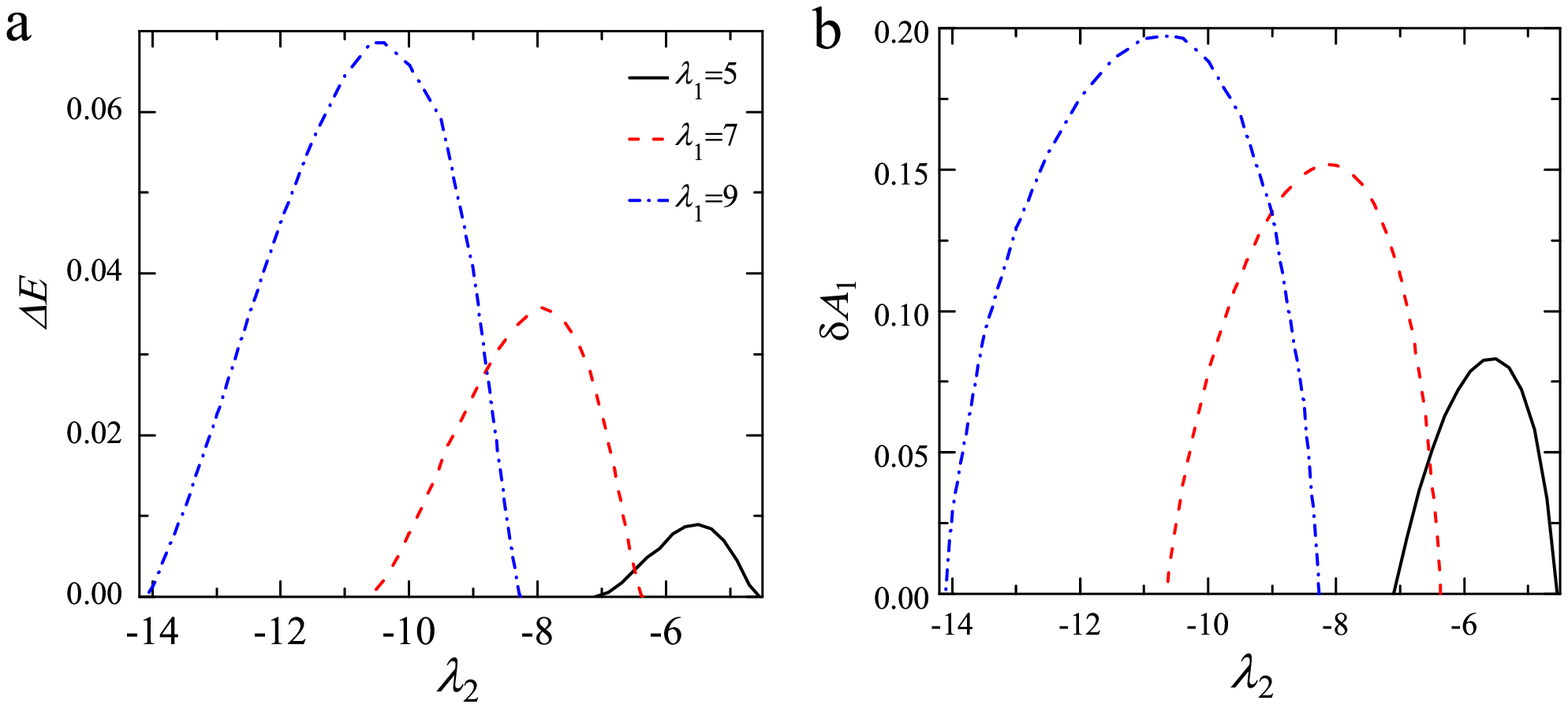}
\caption{\label{Fig6}(a) (Color online)  Energy difference $\Delta
E$ and (b) perturbations $\delta A_1$ for destabilizing the DK as
a function of three-body interactions $\lambda_2$. Different lines
indicate three different values of $\lambda_1$, respectively.
$\lambda_2^*$ is a critical value of $\lambda_2$. When
$\lambda_2=\lambda_2^*$,  the DK is the most stable one for the
fixed $\lambda_1$. $\delta \theta_{12}=0$ in (b).}
\end{figure}

\subsection{The stability and dynamics of DK}\label{sec3.2}

One  can  investigate the  stability and  dynamics of DK with the
same way used above. When $\lambda_2\rightarrow
-\infty$ and  $\lambda_1/\lambda_2 \rightarrow 0$, one can get the
saddle point from Eq. (\ref{eq9}) as
\begin{eqnarray}\label{eq19}
\!\!\!\!A_1\!=\!A_3\approx\!\frac{1}{\sqrt{3}}+\!\frac{\sqrt{3}}{8
\lambda _2}+\!\frac{54 \sqrt{3}}{128 \lambda
_2^2}-\!\frac{3\sqrt{3} \lambda _1}{8 \lambda _2^2}+\!\frac{513
\sqrt{3}}{256 \lambda _2^3}.
\end{eqnarray}
By substituting the saddle point into Eq. (\ref{eq7}) with
$A_1^2+A_2^2+A_3^2=1$, one can get the energy threshold
\begin{eqnarray}\label{eq20}
\!\!E_{\text{thr2}}\!\!\!\!&=&\!\!\!\!\left(\!\!-\frac{\lambda
_2}{27}\!-\!\frac{2}{3}+\frac{1}{8 \lambda _2}\!+\!\frac{29}{64
\lambda _2^2}\!+\!\frac{981}{512 \lambda
_2^3}\!+\!\frac{151119}{16384 \lambda
_2^4}\right) \nonumber\\
&&\!\!\!-\!\!\left(\frac{1}{6}+\frac{3}{16 \lambda
_2^2}\!+\!\frac{45}{32 \lambda _2^3}\!+\!\frac{9369}{1024 \lambda
_2^4}\!+\!\frac{3013119}{32768 \lambda _2^5}\right)\!\! \lambda
_1\nonumber\\ &&\!\! +\!\!\left(\frac{135}{16 \lambda
_2^4}+\frac{243}{8 \lambda _2^5}\right)\!\!\lambda
_1^2\!\!-\!\!\left(\frac{27}{16 \lambda _2^4}+\!\frac{189}{128
\lambda _2^5} \right)\!\!\lambda _1^3.
\end{eqnarray}

Next, we consider the fixed  point corresponding to the DK which
is also gained from Eq. (\ref{eq9}). The initial condition for the
DK reads
$\overrightarrow{\psi}^{DK}(t=0)=(A_1^{DK},A_2^{DK},A_3^{DK})$.
One can add perturbations to site 1:
$\overrightarrow{\psi}{(0)}=((A_1^{DK}+\delta_
{A_1})e^{i\delta_{\theta}},A_2,A_3^{DK})$, where $A_2 =
(1-|\psi_1|^2-|\psi_3|^2)^{1/2}$. Dynamics on the PN landscape for
increasing total energy of the trimer with fixed two-body
interactions $\lambda_1=6$ and three-body interactions
$\lambda_2=-7.5$ is shown in Fig. 5. Here
$E_{thr2}=-1.44281$. If the perturbation is small, $E_t<E_{thr}$,
the areas in phase space are disconnected. The DK is stable and
the norm is localized nearly in the adjacent two sites $2$ and $3$
on short time scales, shown in Fig. 5(a) and (e). If
$E_t$ is just larger than $E_{thr}$, the areas in phase space are
connected and instability of the DK can be observed; that is,  the
DK migrates from site $3$ to site $1$ and then tangles at $1$ or
$3$, but the norm of site $2$ [presented by $A_2(t)$] is nearly
constant, shown in Fig. 5(b) and (f). If the perturbation
becomes larger, one can see that there are long-range and
long-lived Josephson oscillations between sites 1 and 3 with
negligible variation of norm in site 2, shown in Fig.
5(c) and (g). If the perturbation is large enough, the
orbit explores large parts of space and visits all three sites,
shown in Fig. 5(d), and large amplitudes $A_n(t)$ can be
found at all three sites, shown in Fig. 5(h).

One can also use $\delta A_1$ and $\Delta E$ to
investigate the stability of DK, shown in Fig. 6.
For a fixed $\lambda_1$, one can see that
both $\Delta E$ and
$\delta A_1$ first increase  and then decrease with $\lambda_2$.
That is, there exists a critical three-body interaction
$\lambda_2^*$ with which the DK is the most
stable one.  When $\lambda_2 \neq \lambda_2^*$, the DK becomes more
unstable when $\lambda_2$ decreases or increases. Furthermore, we
numerically get the critical value when $\lambda_1>3.5$
\begin{eqnarray}\label{eq22}
\lambda_2^*&=&0.27034-1.13186\lambda _1+0.00516\lambda _1^2.
\end{eqnarray}
It is presented by the dotted line in phase III in Fig. 1.

Obviously,  the stability of  the DK depends strongly on both two-
and three-body interactions, while the DK is more unstable than the
ordinary DB.

\subsection{The stability and dynamics of MUB}\label{sec3.3}

For MUB, the lower PN energy landscape in
Eq. (\ref{eq1a}) becomes
\begin{eqnarray}\label{eq23}
\!\!H_{\text{PN}}^l\!\!&\!\!=\!\!&\!-\!\frac{\lambda_1
}{2}\!\left(\!A_1^{4}\!+\!A_2^{4}\!+\!A_3^{4}\!+\!A_4^{4}\right)\!\!
-\!\!\frac{\lambda_2
}{3}\!\left(\!A_1^{6}\!+\!A_2^{6}\!+\!A_3^{6}\!+\!A_4^{6}\right)\;\;\;\;\nonumber\\&&\!-\left(A_1
A_2+A_2 A_3+A_3 A_4\right).
\end{eqnarray}
When $\lambda_2\rightarrow -\infty$ and  $\lambda_1/\lambda_2
\rightarrow 0$,  one can get the saddle point from Eq. (\ref{eq16})
as
\begin{eqnarray}\label{eq24}
A_1&\!\!\approx\!\!&\frac{1}{\sqrt{3}}+\frac{\sqrt{3}}{8 \lambda
_2}+\frac{54 \sqrt{3}}{128 \lambda _2^2}-\frac{3\sqrt{3} \lambda
_1}{8\lambda
_2^2}+\frac{513\sqrt{3}}{256\lambda _2^3},\nonumber\\
A_3&\!\!\approx\!\!&\frac{1}{\sqrt{3}}+\frac{\sqrt{3}}{8 \lambda
_2}-\frac{54 \sqrt{3}}{128 \lambda _2^2}-\frac{3\sqrt{3} \lambda
_1}{8\lambda _2^2}+\frac{513\sqrt{3}}{256\lambda
_2^3},\nonumber\\A_4&\!\!\approx\!\!&\frac{27 \sqrt{3}}{4\lambda
_2^2}.
\end{eqnarray}
By substituting the saddle point into Eq. (\ref{eq23}) with
$A_1^2+A_2^2+A_3^2+A_4^2=1$, one can get the energy threshold
\begin{eqnarray}\label{eq25}
\!\!\!\!\!\!\!E_{\text{thr3}}\!\!\!\!&=&\!\!\!\!-\frac{\lambda
_2}{27}-\frac{2}{3}+\frac{1}{8 \lambda _2}-\frac{\lambda
_1}{6}-\frac{403+12 \lambda _1}{64 \lambda
_2^2}\nonumber\\&&\!\!\!\!-\frac{3 \left(-8541+3072 \lambda _1+640
\lambda _1^2\right)}{2048 \lambda
_2^3}\nonumber\\&&\!\!\!\!+\frac{9\left(103827\!+\!61338 \lambda
_1\!+\!8032 \lambda _1^2\!-\!3584 \lambda _1^3\right)}{8192
\lambda _2^4}.
\end{eqnarray}

Suppose that the initial condition for the MUB is
$\overrightarrow{\psi}^{MUB}(t=0)=(A_1^{MUB},A_2^{MUB},A_3^{MUB},A_4^{MUB})$.
One can add perturbations to site 1:
$\overrightarrow{\psi}{(0)}=((A_1^{MUB}+\delta_
{A_1})e^{i\delta_{\theta}},A_2,A_3^{MUB},A_4^{MUB})$, where $A_2 =
(1-|\psi_1|^2-|\psi_3|^2-|\psi_4|^2)^{1/2}$. Dynamics of the MUB for
increasing total energy of the four-site system with $\lambda_1=6$
and $\lambda_2=-7$ is shown in Fig. 7. Here
$E_{thr3}=-1.50027$. If there is no perturbation, i.e., $\delta
A_1=\delta \theta=0$, the MUB is stable absolutely and the norm is
localized in the middle adjacent two sites $2$ and $3$ with
$A_2(t)=A_3(t)$ and $A_1(t)=A_4(t)$, shown in  Fig. 7(a).
If the perturbation is small, $E_t<E_{thr3}$,  the MUB is stable and
the norm is localized nearly in the adjacent two middle sites $2$
and $3$, shown in Fig. 7(b). In contrast, if $E_t$ is just
larger than  $E_{thr3}$, the MUB becomes unstable after $t \approx
78$ time steps, and the MUB migrates from sites $2$ and $3$ to sites
$3$ and $4$ or $1$ and $2$, shown Fig. 7(c). Similarly, if
the perturbation is large enough,  the stability of the MUB is
destroyed, shown in Fig. 7(d).

Also, one can use $\delta A_1$ and $\Delta E$ to investigate the stability of MUB, shown in  Fig. 8. For a fixed
$\lambda_1$, both $\Delta E$ and $\delta A_1$ first increase with
$\lambda_2$ before decreasing. That is, there still exists a
critical three-body interaction $\lambda_2^*$ at which the MUB is the most stable one. When $\lambda_2 \neq
\lambda_2^*$, the MUB becomes unstable when $\lambda_2$ decreases or
increases.

\begin{figure}     
\includegraphics[width=12cm]{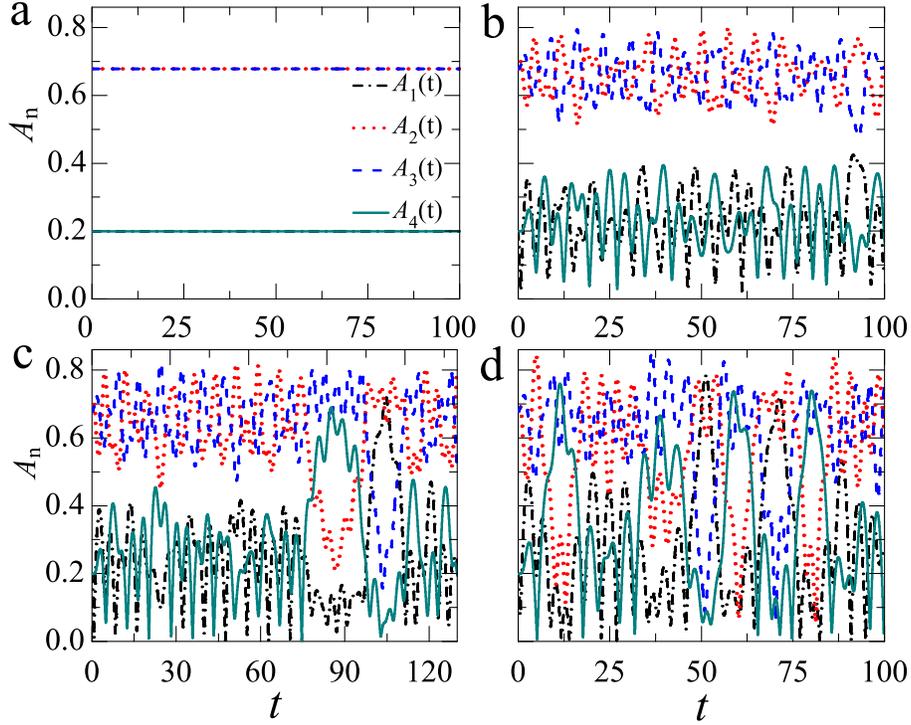}
\caption{\label{Fig7} (Color online) Dynamics of MUB for
increasing the total energy of the four-site system with
$\lambda_1=6$ and $\lambda_2=-7$, where $E_{thr3}=-1.50027$. (a)
$\delta_A=\delta_{\theta}=0$, the MUB is absolutely stable, and
$A_1(t)=A_4(t)$ and $A_2(t)=A_3(t)$. (b)
$\delta_{\theta}=\pi/3.4$, $E_t=-1.50212<E_{thr3}$, the
multi-breather is stable, and it is still located at sites $2$ and
$3$. (c) $\delta_{\theta}=\pi/2.8$, $E_t=-1.47925$ is just larger
than $E_{thr3}$, the multi-breather is unstable, and the norm from
sites $2$ and $3$ migrate to sites $3$ and $4$ after $t \approx
78$ time steps. (d) $\delta_\theta=\pi/2$,
$E_t=-1.42046>E_{thr3}$, the stability of the multi-breather is
destroyed, and the dominating amplitude can be found in any of the
four sites. In all cases $\delta_{A_1}=0$. Initial conditions are
given by Eq. (\ref{eq16}).}
\end{figure}

\begin{figure}    
\includegraphics[width=12cm]{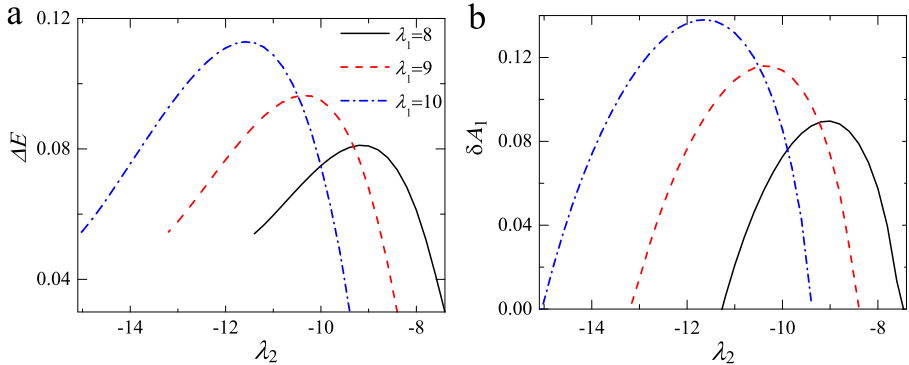}
\caption{\label{Fig8} (Color online) (a) Energy difference $\Delta
E$ and (b) perturbations $\delta A_1$ for destabilizing the MUB as
a function of three-body interactions $\lambda_2$. Different lines
indicate three different values of $\lambda_1$, respectively. The
$\lambda_2^*$ is a critical value of $\lambda_2$. When
$\lambda_2=\lambda_2^*$, the MUB is the most stable for the fixed
$\lambda_1$. $\delta \theta_{12}=\pi/3.4$ in (b).}
\end{figure}

\section{Extended lattices}\label{sec4}

Now, let us generalize our study to the case with   extended
lattices, i.e., $M>4$.

\subsection{DB in extended lattices}\label{sec4.1}

Here we use the same initial conditions
as the case with three sites, i.e., $\lambda_1 = 6$, $\lambda_2=-1.5$, to
study an extended lattice system. We assume $M=101$ and the
DB locates at the site $n=51$. The initial condition
reads
\begin{equation}\label{eq26}
\begin{split}
\psi_{50}(0)&=(A_1^{DB}+\delta_{A_1})e^{i\delta_\theta},\\
\psi_{51}(0)&=A_2,\\
\psi_{52}(0)&=A_1^{DB},\\
\psi_n(0)&=0,\;\; \texttt{else},
\end{split}
\end{equation}
where $A_2=(1-|\psi_{50}|^2-|\psi_{52}|^2)^{1/2}$ and $A_1^{DB}$
is obtained exactly from Eq. (\ref{eq9}). The wave function is
normalized to $\sum_{n=1}^{M}|\psi_n|^2=1$. The stability and
migration of a DB in extended lattices ($M = 101$ sites) is shown
in Fig. 9. The solution for the DB in the trimer, i.e.,
the initial condition shown in Eq.(\ref{eq26}), is inserted in the
extended lattice. Here $E_{thr1}=-1.91469$ when $\lambda_1=6$ and
$\lambda_2=-1.5$ (corresponding to Fig. 3). Obviously,
when no perturbation is added to the site $50$, i.e.,
$\delta_{A_1}=\delta_{\theta}=0$, the breather is stable, shown in
Fig. 9(a) and (d). When the perturbation is added to the
site $50$ and the total energy of the local trimer $E_t$ is just
larger than $ E_{thr1}$,  the breather is  stable and no migration
takes place, different from the case with three sites,  shown in
Fig. 9(b) and (e). The reason is that the energy can flow
into additional degrees of freedom in extended lattices. When the
perturbation is large and the total energy of the local trimer
$E_t> E_{thr1}$,  the breather is destabilized and migrates from
site $51$ to site $50$ after $t \approx 2$ time steps, similar to
the case with the three-site model, shown in Fig. 9(c)
and (f).

\begin{figure*}    
\includegraphics[width=16cm]{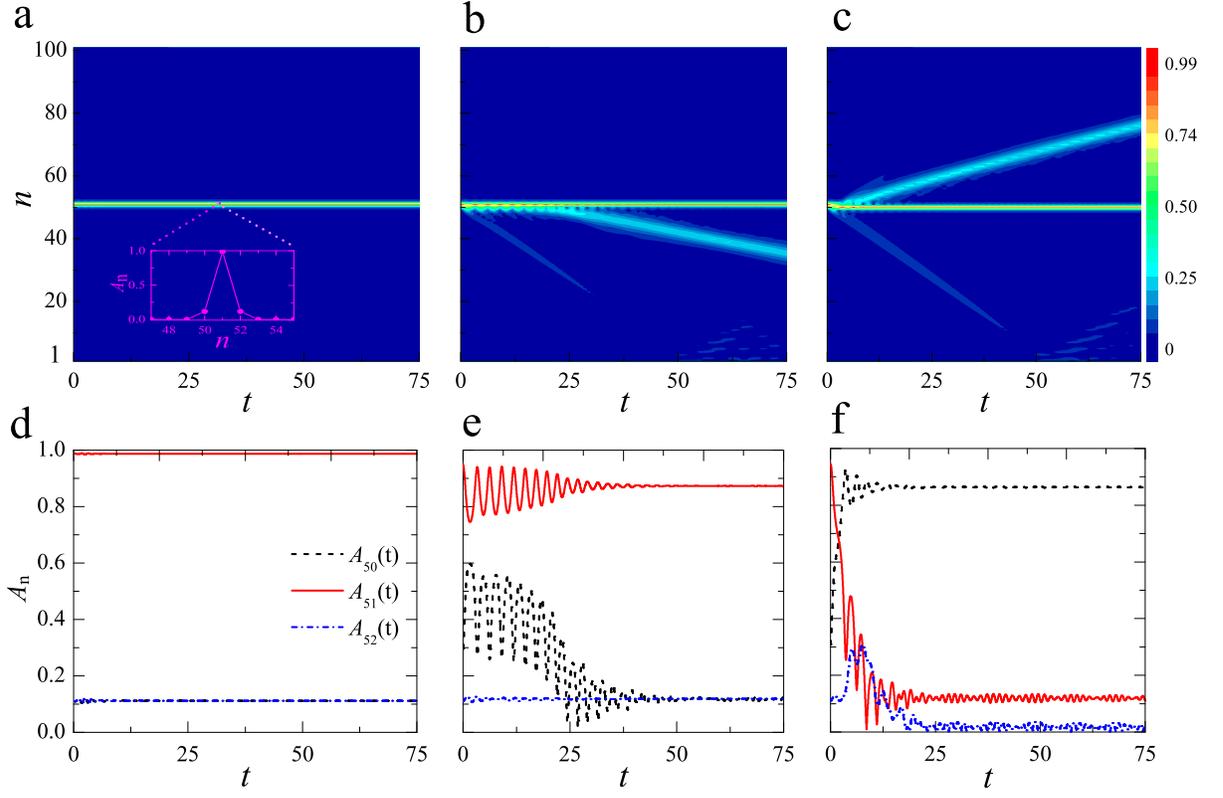}
\caption{\label{Fig9} (Color online) The stability and migration
of a DB in extended lattices ($M = 101$ sites). The color code
represents $|\psi_n(t)|$. The insert in (a) represents the
structure of DB. Initial conditions are given by Eq. (\ref{eq26})
with (a) $\delta_A=\delta_\theta=0$, $E_t=-2.61029<E_{thr1}$, (b)
$\delta_A=0.184$, $\delta_\theta=\pi$, $E_t=-1.91179$ is just
larger than $E_{thr1}$, (c) $\delta_A=0.199$, $\delta_\theta=\pi$,
$E_t=-1.87136> E_{thr1}$. In all cases $\lambda_1=6$ and
$\lambda_2=-1.5$.}
\end{figure*}

\begin{figure*}     
\includegraphics[width=16cm]{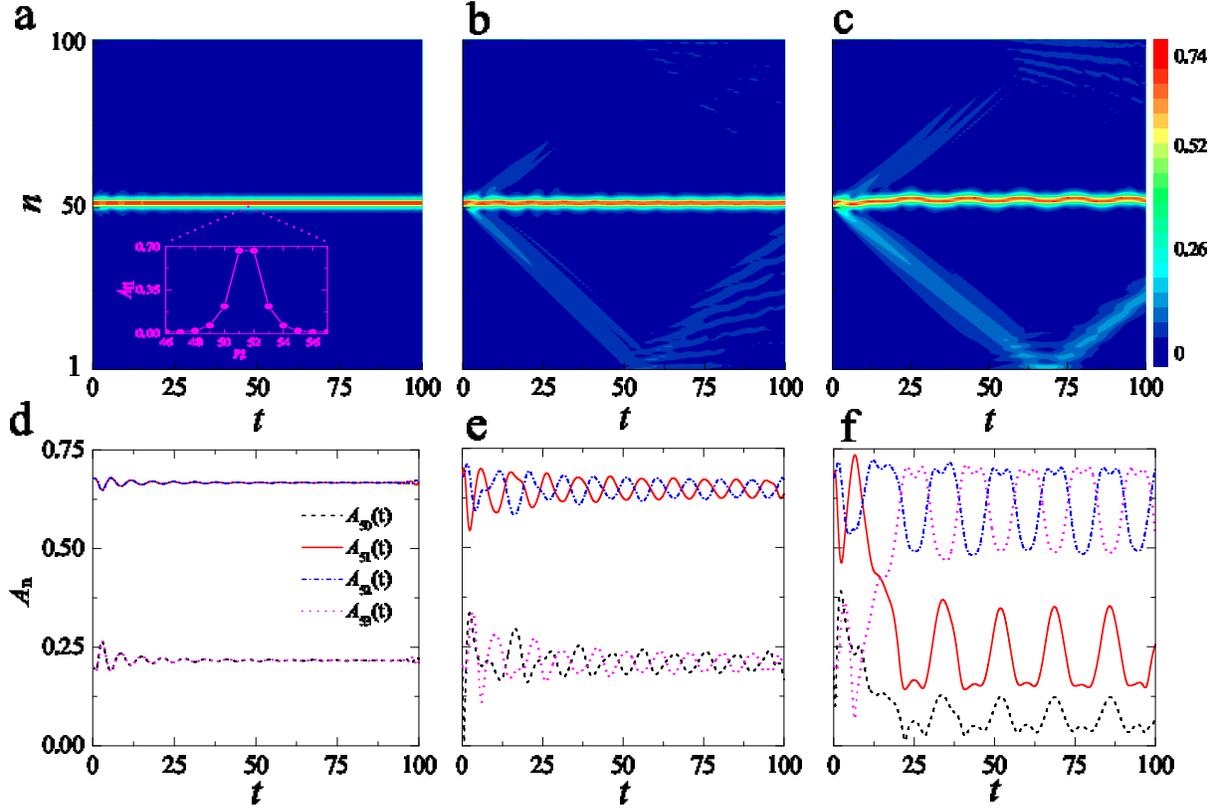}
\caption{\label{Fig10} (Color online) The stability and migration
of a MUB in extended lattices ($M = 101$ sites). The color code
represents   $|\psi_n(t)|$. The insert in (a) represents the
structure of MUB. Initial conditions are given by Eq. (\ref{eq16})
with (a) $\delta_A=\delta_\theta=0$, $E_t=-1.55596<E_{thr3}$, (b)
$\delta_\theta=\pi/2.8$, $E_t=-1.47925$ is just larger than  $
E_{thr1}$, (c) $\delta_\theta=\pi/1.79$, $E_t=-1.39563> E_{thr3}$.
In all cases $\lambda_1=6$ and $\lambda_2=-7$.}
\end{figure*}

\subsection{MUB in extended lattices}\label{sec4.2}

For MUB, we use the same initial condition as the case with the
four-site model to study its stability and dynamics in the extended
lattice system (M=101 sites), i.e., $\lambda_1 = 6$ and
$\lambda_2=-7$. Here, the MUB is located at the sites $n=51$ and
$n=52$. The initial condition reads
\begin{equation}\label{eq27}
\begin{split}
\psi_{50}(0)&=(A_1^{MUB}+\delta_{A_1})e^{i\delta_\theta},\\
\psi_{51}(0)&=A_2,\\
\psi_{52}(0)&=A_2^{MUB}, \\
\psi_{53}(0)&= A_1^{MUB}, \\ \psi_n(0)&= 0,\;\; \texttt{else},
\end{split}
\end{equation}
where $A_2=(1-|\psi_{50}|^2-|\psi_{52}|^2-|\psi_{53}|^2)^{1/2}$
and $A_1^{MUB}$ and $A_2^{MUB}$ are obtained exactly from Eq.
(\ref{eq16}). The wave function is normalized to
$\sum_{n=1}^{M}|\psi_n|^2=1$. The solution for the MUB in the
four-site model  is inserted in the extended lattice. Here
$E_{thr3}=-1.50027$ when $\lambda_1=6$ and $\lambda_2=-7$.
Obviously, when no perturbation is added to site $50$, i.e.,
$\delta_{A_1}=\delta_{\theta}=0$,  the breather is stable, shown
in Fig. 10(a) and  (d). As
$|\psi_{50}(t)|=|\psi_{53}(t)|$ and
$|\psi_{51}(t)|=|\psi_{52}(t)|$,  the red and the blue lines are
overlapped, and the black and the green lines are overlapped in
Fig. 10(d). When the perturbation is added to  site $50$
and the total energy $E_t$ is just larger than $ E_{thr3}$, the
energy can flow into additional degrees of freedom in extended
lattices, and the MUB is still stable and no migration takes
place, shown in Fig. 10(b) and  (e).  When the
perturbation is larger and the total energy $E_t> E_{thr3}$,  the
MUB is destabilized  and migrates from site $51$ to site $53$ and
locate in sites $52$ and $53$ after $t \approx 14$ time steps,
shown in Fig. 10(c) and  (f).

From the discussion above, one can see that the stability and the dynamics characters of both DB and MUB are general for  extended lattices.

\section{Discussion and summary}\label{sec5}

Although the three-body interactions could be observed or
realized in experiment and theory \cite{SWill2010,DaleyAJ2014}, there are no systematical
analysis of their type,  existence, and stability of the localized
modes. Previous works indicated that DB can exist and its stability
plays  a crucial role in blocking, filtering, and transfer in BECs
\cite{GPTsironis1996,RLivi2006,GSNg2009,BRumpf2004}. The exact
condition to destabilize DB is still not clear, especially in the
case by considering three-body interactions. Recent works
\cite{GSNg2009,RFranzosi2011} showed that MUB can exist for BECs in
optical lattices by only considering two-body interactions in the
case with large interactions and appropriate phases.

In our work, we systematically investigated the existence of the
different localized modes for different parameters $(\lambda_1,
\lambda_2)$, i.e.,  DB, DK, and MUB, and discussed the effect of
two- and three-body interactions on their stabilities. Both on-site
repulsive two- and three-body interactions can stabilize DB, while
on-site attractive three-body interactions destabilize DB. DK and
MUB are the most stable ones when the three-body interactions are
equal to a critical value for the fixed two-body interactions. It
may provide effective guidance to gain different kinds of localized
modes in optical lattices \cite{HHennig2013,RLivi2006} and help us
to study its properties in experiment, especially for MUB.

Besides, in our work, three analyzed thresholds to destabilize the
three localized modes are given explicitly. If the total energy of
the system is higher than the energy thresholds, the localized
state is unstable. On the contrary, if the total energy of the
system is lower than the energy thresholds, the localized state is
stable. It may lead to some interesting applications for blocking
and filtering atom beams when there are both two- and three-body
interactions in the system. Furthermore, it is useful
for controlling the transmission of matter waves in interferometry
and quantum-information processes \cite{RAVicencio2007}.

In summary, we have investigated the stability and phase transition of
localized modes in BECs in an optical lattice with the discrete
nonlinear Schr\"{o}dinger model by considering two- and three-body
interactions. It has been shown that there are three different types
of localized modes. The first one is bright DB which can be
stabilized by both on-site repulsive two- and three-body
interactions. However, on-site attractive three-body interactions
destabilize DB. The second one is DK which  is the most stable one
when the three-body interaction is equal to a critical value for
fixed two-body interactions.  The third one is MUB. It becomes the
most stable one when three-body interaction is in the critical
value. Moreover, the stability and dynamics characters of DB and MUB
are general for  extended lattice systems.
 Our results   provide a deep insight into the
dynamics of blocking, filtering, and transfer of the norm in
nonlinear lattices for BECs by considering both two- and
three-body interactions.

\section*{Acknowledgments}

This work is supported by the National Natural Science Foundation of
China under Grant Nos.  11174040, 11475021, and 11474027, and
NECT-11-0031.

\end{document}